\documentstyle[12pt, aaspp4]{article}

\newcommand{\lsun}{log $L/L_{\odot}\,$}
\newcommand{\msun}{$M/M_{\odot}\,$}

\begin{document}

\title{On the Theoretical Period-Radius Relation of Classical Cepheids}

\author{Giuseppe Bono\altaffilmark{1}, Filippina Caputo\altaffilmark{2}, 
and Marcella Marconi\altaffilmark{3}}

\lefthead{Bono et al.}
\righthead{Cepheid PR relation} 

\altaffiltext{1}{Osservatorio Astronomico di Trieste, Via G. B. Tiepolo 11, 
34131 Trieste, Italy; bono@oat.ts.astro.it}
\altaffiltext{2}{Osservatorio Astronomico di Capodimonte, Via Moiariello 16,
80131 Napoli, Italy; caputo@astrna.na.astro.it} 
\altaffiltext{3}{Dipartimento di Fisica, Univ. di Pisa, Piazza Torricelli 2,
56126 Pisa, Italy; marcella@astr1pi.difi.unipi.it}

\begin{abstract}

We present the results of a comprehensive theoretical investigation 
on the Period-Radius (PR) relation of classical Cepheids based 
on new sequences of full amplitude, nonlinear convective models 
constructed by adopting a wide range of both stellar masses and 
chemical compositions. In the period range $ 0.9 \le log P \le 1.8$ 
a very good agreement is found between theoretical predictions and 
current available data whereas outside this range, both at shorter 
and at longer periods, nonlinear radii attain intermediate values 
between empirical relations based on different Baade-Wesselink (BW)
methods and photometric bandpasses.  

\end{abstract}

\noindent
{\em Subject headings:} Cepheids --- galaxies: stellar content --- 
stars: fundamental parameters  --- stars: oscillations

\pagebreak
\section{Introduction}

The BW method (Baade 1926; Wesselink 1946) has been 
receiving growing attention from the astronomical community since it 
allows direct measurement of both radii and absolute magnitudes. 
Even though some physical assumptions of this method were questioned 
(Karp 1975; Gautschy 1987; Bono, Caputo \& Stellingwerf 1994;
Butler et al. 1996), in the last years a paramount 
effort has been undertaken for improving its accuracy and consistency 
(Barnes \& Evans 1976; Sollazzo et al. 1981; Laney \& Stobie 1995, 
hereinafter LS; Ripepi et al. 1997, hereinafter RBMR). At the same time, 
Krockenberger, Sasselov \& Noyes (1997, hereinafter KSN) have recently 
developed a new BW method, based on Fourier coefficients, for evaluating 
the uncertainty on mean stellar radii due to individual measurement errors. 

Substantial improvements in the measurements of both Cepheid radii 
and distances were thoroughly discussed in several outstanding 
papers by LS and more recently by Laney (1997), Di Benedetto (1997) 
and Gieren, Fouqu\`e \& G\`omez (1997, hereinafter GFG). 
Despite this ongoing observational effort, theoretical investigations 
devoted to the Cepheid PR relation based on up-to-date evolutionary and
pulsational models are lagging. In fact LS, by comparing the PR 
relation derived for a sample of 49 Galactic Cepheids with Fernie's (1984) 
weighted mean theoretical PR relation, found that the slope of the 
empirical relation is steeper than the theoretical one and that BW radii 
are 12\% smaller than the theoretical ones for a period equal to 10 d. 
On the other hand RBMR, by adopting a new version of the CORS method
(Sollazzo et al. 1981, and references therein) found, as expected (LS), 
that the slope of their PR
relation is slightly shallower if compared either with empirical BW
relations based on IR photometry, or with theoretical relations.

The reason why so far only few investigations have been devoted to the 
evaluation of the mean theoretical PR relation is that its slope  
depends on the intrinsic width of the instability strip. The cool edge 
of the instability strip can be evaluated only by coupling the local 
conservation equations with a nonlocal and time-dependent equation 
for turbulent-convective motions (Stellingwerf 1982; Gehmeyr 1992). 
Theoretical PR relations available in the literature (Karp 1975; 
Cogan 1978) are based on radiative models and therefore cannot be 
considered "pure" theoretical relations.  
In fact, radiative models can only fix the location of the blue edge, 
whereas the temperature width of the instability strip is inferred 
from observational data. As a consequence, both the zero-point and 
the slope of these "semi-theoretical" PR relations depend on the 
completeness of the adopted sample and on the relations used for 
transforming the mean colors into mean effective temperatures.  
Moreover, Karp's and Cogan's relations have been derived by assuming 
that the width 
of the instability strip is constant when moving from short to 
long-period Cepheids. However, this assumption is not supported 
by observational  estimates, and indeed Pel (1980) in a seminal 
investigation showed that the Cepheid instability region is not a 
rectangular-shaped but a wedge-shaped strip, i.e. the color range 
narrows toward short-period Cepheids. 
The main aim of this investigation is to establish the Cepheid PR 
relation on a genuine theoretical basis by adopting the mean radii 
and the periods predicted by full amplitude, nonlinear convective models
and then to compare theoretical with empirical PR relations. 

\section{Pulsational Models}

Several sequences of envelope models were constructed by adopting four 
different stellar masses (\msun=5.0, 7.0, 9.0, 11.0) and two luminosity 
levels for each mass value. 
The luminosity levels were fixed according to the Mass-Luminosity (ML)  
relations predicted both by canonical (no overshooting) and noncanonical 
(mild overshooting, $\lambda_{over}$=0.5) evolutionary models. The former 
relation was chosen from the calculations of Castellani, Chieffi \& 
Straniero (1992), whereas the latter was fixed by increasing the canonical 
luminosity level by 0.25, i.e. \lsun$(NC) =$ \lsun$(C) + 0.25$ 
(see e.g. Chiosi et al. 1992 and Chiosi, Wood \& Capitanio 1993).  
Our investigation is also focussed on the dependence of the PR relation 
on He and metal contents, and therefore calculations were performed 
by adopting three different chemical compositions which are representative
of Cepheids in the Small (Y=0.25, Z=0.004), the Large (Y=0.25, Z=0.008)
Magellanic Cloud (MC), and in the Galaxy (Y=0.28, Z=0.02). 
The models were arranged in sequences characterized by constant mass, 
luminosity and chemical composition but by different values of the 
effective temperature ($4000 \le T_e \le 7000$ K). 
Physical and numerical assumptions adopted for performing the linear,
nonadiabatic analysis as well as the nonlinear, full amplitude analysis
have already been the subjects of previous papers (Bono \& Stellingwerf 1994; 
Bono \& Marconi 1997, and references therein), and therefore they are not 
further discussed here.  
The theoretical framework we developed proved to be successful in 
reproducing observational properties (amplitudes and modal stability) 
of Cepheids characterized by periods shorter than 40 d. 
This notwithstanding, we found that high-mass models -\msun=11.0- present 
peculiarities in the nonlinear limit cycle stability.  
In fact, light and velocity curves display irregularities such as sharp 
bumps and sudden dips during both contraction and expansion phases. 
Moreover, pulsational properties undergo substantial changes 
over consecutive periods. A similar behavior was found by Christy 
(1975), who only pointed out that both pulsation irregularities and very 
large amplitudes take place in models with high radius/mass, period/radius, 
and period/luminosity ratios.  

For investigating the intimate nature of this phenomenon a detailed 
analysis of the dependence of the limiting amplitude behavior on 
physical and/or numerical assumptions was undertaken. 
We found that pulsation irregularities are caused by 
the coarse spatial resolution in the H and first He (HeI) 
ionization regions. In fact these layers, due to their large 
back and forth motion, over a full cycle undergo a large excursion 
both in temperature and density. The coarse spatial 
resolution causes a sudden increase in the temperature and density 
gradients and consequently the formation and propagation of strong 
spurious shocks. This is a typical limit of the Lagrangian models 
when compared with the adaptive grid models. For solving this 
problem we constructed a new sequence of linear models. 
The main differences between these equilibrium models and the standard
ones are the following:

standard models are constructed, following Stellingwerf (1975), by 
anchoring the opacity peak of the H ionization regions (HIR) and by 
locating a proper number of zones ($20\div30$) between this peak 
and the surface layer. 
Instead of improving accuracy by simply increasing the number 
of zones located above this peak we developed a new method which, by 
means of a multiple iteration on the mass of the surface zone, ensures 
via a secant method a uniform sampling ($\Delta T = 500\div650$ K) 
of the layers located between the surface and the base of the H and 
HeI ionization regions ($T \approx 2.1\times 10^4$ K). 
The left and the right panels of Figure 1 show the opacity and the 
adiabatic exponent $\Gamma_1$ of two models located close to the blue 
and to the red edge of the instability strip. The fine models  
show two substantial differences when compared to the coarse  
ones: {\em a}) {\em their adiabatic index attains smaller values and 
resolve the HeI ionization zone}. 
The HeI region is dynamically unstable, and 
indeed close to $log T\approx4.2$  the $\Gamma_1$ is smaller than 4/3. 
{\em b}) {\em Their opacity peak attains larger values}, the differences 
are a factor of four in the blue models and of the order of 10\% in 
the red models. The same differences were found by Gehmeyr (1992) in 
his comparison of two static RR Lyrae models constructed by adopting a 
Lagrangian and an adaptive grid code respectively.  
It is hardly necessary to point out the role played by these changes 
in the instability and pulsation amplitudes of long-period Cepheids.  

The nonlinear radii and periods discussed in this investigation were 
evaluated by adopting fine zoning models. 
Figure 2 shows the comparison between the theoretical PR relations at 
solar metallicity provided by Cogan (1978) and Karp (1975) and our 
models constructed by adopting canonical (triangles) and noncanonical 
(squares) ML relations. The nonlinear radii of canonical Cepheids 
are larger than the noncanonical ones, with a difference ranging from 
4\% at $log P\approx$0.6 to 7\% at $log P\approx$0.6. Interesting enough, 
our nonlinear radii are quite similar to the radii predicted by Cogan's 
relation in the range $0.6 \le log P \le 1.2$, whereas toward longer 
periods they first attain values similar to those given by Karp's relation 
and then become systematically smaller than the radii predicted by the quoted 
relations. 
This difference is mainly due to the proper location of red boundaries 
without invoking ad hoc assumptions and, more marginally, to new opacities. 
Table 1 summarizes the zero points and the slopes of the linear regression
obtained by adopting different compositions and ML relations. An interesting 
result is that the average PR relations show a mild but non 
negligible dependence on metal content. In fact, for canonical radii an 
increase in the metal content from Z=0.004 to Z=0.008 leads to a decrease 
which ranges from 2\% at $log P\approx$0.6 to almost 4\% at 
$log P\approx$2.0. An increase in both He and metal contents 
(Y=0.28, Z=0.02 against Y=0.25, Z=0.004) implies a decrease which 
ranges from 4\% at $log P\approx$0.6 to 9\% at $log P\approx$2.0.  
A similar outcome results for noncanonical radii. 

\section{Comparison Between Theory and Observations}

Figure 3 shows the empirical PR relations for Galactic Cepheids 
obtained by GFG (dashed line), Laney (1997, long-dashed lines), and 
CORS (dotted line). Theory and observations were also compared by  
plotting canonical periods and radii of models with Z=0.008 and Z=0.02. 
We adopted two different compositions for accounting for the spread in 
metal content recently found by Fry \& Carney (1997) among calibrating 
Galactic Cepheids. The comparison brings out two major results: 
{\em a}) theoretical predictions are, within the observational errors, 
in good agreement with average empirical PR relations obtained by adopting 
different methods and different photometric bands. In the period range 
$0.9 \le log P \le 1.8$ observed and theoretical radii are almost identical.  
{\em b}) Theoretical radii are systematically larger than the observed ones 
in the period range $0.4 \le log P \le 0.6$, whereas they are smaller 
toward longer periods, i.e. $log P > 1.8$.  
Both the paucity of long-period Cepheids detected in the Galaxy and the 
lack of a detailed analysis of the systematic errors involved
in empirical PR relations based on different methods and/or photometric
bands prevent a quantitative explanation of this discrepancy. 

However, KSN have recently provided a thorough analysis of the uncertainty
in the radius estimates introduced by individual measurement errors.    
Their results on the slope of the PR relation for Galactic Cepheids fairly
agree with those of RBMR who adopted a CORS method which accounts for the 
loop performed by the variable in a color-color plane, the main advantage 
of this method being its independence of reddening corrections. 
Moreover, Di Benedetto (1997) obtained a very precise general PR relation 
by adopting both Galactic and MCs Cepheids for which high precision 
photometric and spectroscopic data were available. 
In this method the use of both magnitude and colors in evaluating stellar 
angular sizes ensures a marginal dependence of radii on both reddening and 
metallicity. 

Figure 4 shows the last two empirical PR relations, the results obtained 
by KSN as well as the theoretical predictions for the three chemical 
compositions. 
The comparison discloses once again a remarkable agreement between theory 
and observations. The major discrepancy is in the short-period range,  
in which theoretical radii are smaller than the radii obtained by KSN 
and RBMR, and larger than the radii provided by Di Benedetto's relation. 
However, firm constraints on this observational discrepancy cannot be 
drawn since, as KSN clearly stated, the uncertainty in the mean radii 
are dominated by the error in the phase difference between color index 
and magnitude. Moving toward short-period Cepheids this difference 
becomes smaller, and in turn the uncertainty becomes larger. 
This trend is reversed in the long-period range, and indeed
for periods longer than 30 d the radii obtained by KSN and RBMR are
systematically smaller than the estimates of other authors,
theoretical radii being located once again between these two different
estimates. This finding confirms the results obtained by LS concerning 
the systematic error which affects radius estimates, i.e. by neglecting
the variation of the effective gravity over the pulsation cycle, the 
radii based on optical bands systematically underestimate (overestimate) 
the radii of long (short) period Cepheids. 
Since Cepheid radii are proportional to the {\em p}-factor, i.e. the 
factor adopted for converting observed radial velocities into pulsational 
velocities, we suspect that this parameter is not only phase-dependent 
and that its value depends on both the BW method and the data sets adopted 
for estimating the radii (see e.g. KSN), but also that it should attain 
smaller values in long-period Cepheids observed in the IR bands.  
In fact, data in Figure 4 suggest that the dependence of {\em p} on period 
is stronger than predicted by Gieren, Barnes \& Moffett (1989) relation.   

\section{Conclusions}

We developed a new theoretical scenario of the actual properties of 
classical Cepheids in the Galaxy and in the MCs. 
By adopting both radii and periods predicted by full amplitude, nonlinear,  
convective models we found that the use of two different ML relations 
based on canonical and noncanonical (mild overshooting) evolutionary 
models has a marginal effect on the PR relation, and indeed in the mean 
PR relation the difference is of the order of 3\%. 
At the same time, we also found that an increase in the metal content 
implies a decrease in the mean radius. This effect is not constant but 
increases when moving from short to long-period Cepheids. 
In particular, a change in the chemical composition from Y=0.25, Z=0.004
to Y=0.28, Z=0.02 implies at $log P\approx2$ a decrease in the mean radius 
of the order of 9\%. This result prompts that, within the current accuracy 
of both photometric and spectroscopic data, the dependence of the PR relation 
on metallicity could be detected and measured if a proper number of 
long-period variables -$P>40$ d- are included in the sample.  

Theoretical and empirical radii are found in very good agreement in the 
period range $0.9 \le log P \le 1.8$, but present some discrepancies toward 
short and long-period Cepheids. 
No firm conclusion was reached on the intimate nature of this discrepancy
since current mean stellar radii estimated by adopting different BW methods, 
photometric bands, and data sets present a large scatter both at 
$log P < 0.7$ and $log P > 1.8$.  
Comparison between theory and observations suggests that the value 
of the {\em p}-factor could change when moving from short to long-period 
Cepheids. 
At the same time, the results of this investigation disclose a new approach
for testing the internal accuracy and the consistency of the assumptions
adopted by the different BW methods. In fact, observables  
predicted by nonlinear, convective  models can be fed to the progeny 
of the BW method for assessing the intervening effects of systematic 
errors and/or of possible biases in the radius measurements. 

It is a pleasure to thank D. Laney as a referee for his clarifying 
comments and valuable suggestions on current observational data. 

\pagebreak

\pagebreak

\figcaption{Opacity (top) and adiabatic exponent (bottom) as a function 
of the logarithmic temperature for two models located close to 
the blue (left) and to the red (right) edge of the instability strip.  
Solid and dashed lines refer to linear models with fine and coarse spatial 
resolutions in the H and HeI ionization regions, respectively. 
The dotted lines plotted in the bottom panels display the edge between 
dynamically stable ($\Gamma_1 > 4/3$) and unstable ($\Gamma_1 \le 4/3$) 
regions. The arrows mark the main features of the opacity and of the 
adiabatic exponent.} 

\figcaption{Comparison between different theoretical PR relations at 
solar metallicity. Triangles and squares show the nonlinear radii 
obtained by adopting a canonical and a noncanonical ML relation, 
respectively. The dashed line refers to the PR relation obtained by 
Cogan (1978), while the long-dashed line refers to the PR relation provided 
by Karp (1975). Fernie's relation has not been plotted here since it is 
almost identical to Cogan's.} 

\figcaption{Comparison between current empirical PR relations for Galactic 
Cepheids and theoretical nonlinear radii obtained by adopting two different 
chemical compositions.}
 
\figcaption{Comparison between different empirical PR relations and 
theoretical results. Solid and dashed line show Di Benedetto's and RBMR's 
relations, respectively. The former is based on both Galactic and MCs 
Cepheids, whereas the latter on Galactic Cepheids only. Open circles 
refer to the mean radii for Galactic Cepheids obtained by KSN on the basis
of visual magnitudes and B-V colors.} 

\clearpage
\begin{deluxetable}{cccc}
\tablewidth{0pt}
\tablecaption{Theoretical PR Relations  ($log R = \alpha + \beta log P$).}  
\tablehead{ \colhead{Z\tablenotemark{a}} & 
\colhead{$\alpha$\tablenotemark{b}} &
\colhead{$\beta$\tablenotemark{c}}  &  
\colhead{$r$\tablenotemark{d}} }
\startdata
      \multicolumn{4}{c}{Canonical} \nl 
0.02  & 1.188$\pm$0.008\tablenotemark{e} & 0.655$\pm$0.006 & 0.999 \nl  
0.008 & 1.192$\pm$0.009 & 0.666$\pm$0.007 & 0.998\nl  
0.004 & 1.199$\pm$0.010 & 0.670$\pm$0.008 & 0.998\nl 
      \multicolumn{4}{c}{Noncanonical} \nl 
0.02  & 1.174$\pm$0.009 & 0.647$\pm$0.006 & 0.999 \nl 
0.008 & 1.183$\pm$0.009 & 0.653$\pm$0.006 & 0.999 \nl 
0.004 & 1.183$\pm$0.009 & 0.661$\pm$0.006 & 0.998 \nl 
\enddata 
\tablenotetext{a}{ Metallicity.  
\hspace*{0.5mm} $^b$ Zero Points of the PR relations.
\hspace*{0.5mm} $^c$ Slopes of the PR relations.
\hspace*{0.5mm} $^d$ Correlation coefficients of the linear regression.
\hspace*{0.5mm} $^e$ The errors refer to the intrinsic dispersion.}
\end{deluxetable} 
\end{document}